\documentclass[journal]{IEEEtran}

\usepackage{enumerate}
\usepackage{graphicx}
\usepackage{epsf}
\usepackage{subfigure}
\usepackage{amsmath}
\usepackage{amssymb}
\usepackage{array}
\usepackage{setspace}
\usepackage[amsmath,thmmarks]{ntheorem}
\usepackage[ruled,lined]{algorithm2e}
\usepackage{algorithmic}
\usepackage{color}
\usepackage{amsfonts}
\usepackage{dsfont}
\usepackage{xfrac}
\usepackage{breqn}
\usepackage{bbm}
\usepackage{cite}

\graphicspath{{Fig/},{fig/}}

\theoremseparator{.}

\begin{document}

\title{\LARGE Liquid Intelligent Metasurface for Fluid Antennas-Assisted Networks}


\author{
Li-Hsiang Shen,~\IEEEmembership{Member,~IEEE}}

\maketitle

\begin{abstract}
This paper proposes a novel liquid intelligent metasurface (LIM)-assisted downlink multi-user multiple-input single-output (MISO) system, wherein both the base station (BS) and the metasurface are respectively equipped with fluid antennas (FA) and liquid elements. Unlike conventional reconfigurable metasurface-assisted systems with static geometries, the proposed architecture enables joint electromagnetic and spatial reconfigurability by allowing both the FA-empowered BS (FAS) and LIM to dynamically adjust their small-scale positions in addition to beamforming and phase-shift controls. We formulate a sum-rate maximization problem that jointly optimizes the BS beamforming, LIM phase-shifts, and the positions of fluid antennas and liquid elements. The problem is highly non-convex due to coupling between variables, fractional expressions, unit-modulus constraints as well as spatial correlation functions. To address these challenges, we adopt alternating optimization and introduce auxiliary variables and employ Lagrangian dual and quadratic transformation, successive convex approximation (SCA) as well as the penalty convex-concave procedure (PCCP) to solve the respective subproblems. Simulation results have demonstrated that the proposed FAS-LIM architecture significantly outperforms benchmark methods employing conventional fixed metasurface and fixed antenna arrays in terms of various parameter settings.
\end{abstract}


%

{\let\thefootnote\relax\footnotetext
{Li-Hsiang Shen is with the Department of Communication Engineering, National Central University, Taoyuan 320317, Taiwan. (email: shen@ncu.edu.tw)}}

\section{Introduction}

Reconfigurable intelligent surfaces (RISs) have emerged as a transformative paradigm shift in the sixth-generation (6G) wireless communications, enabling the reconfigurable radio environments in a cost-effective and energy-efficient manner \cite{acm, 3, 6}. By leveraging large arrays of low-cost passive elements capable of controlling phase-shifts, RISs can dynamically manipulate electromagnetic waves to enhance signal strength, extend coverage, mitigate interference, and enable robust non-line-of-sight (NLoS) connectivity. This capability has brought up extensive research interest across various domains, including physical layer, network-level spectral and energy efficiency, and intelligent beamforming \cite{2,7, dstar}. However, most existing RIS implementations remain spatially static, limiting their adaptability to dynamic propagation environments and spatial user distributions.

Therefore, the concept of fluid antenna (FA) systems in \cite{9, 10,11} has garnered increasing attention for their ability to dynamically adjust antenna positions, providing spatial diversity and robustness against channel fading, blockage, and interference. By allowing antenna elements to move within a predefined region, FA offers an additional degree of freedom for optimizing wireless links \cite{10, myfas}, particularly in environments with high spatial variability, such as vehicular networks, indoor deployments and dense urban areas \cite{MAA}. While prior studies have explored the use of FA at mobile terminals or base stations (BS), their potential integration with passive RIS remains unexplored.

Motivated by these advancements, this paper proposes a novel FA and liquid intelligent metasurface (LIM) architecture, wherein both the FA-empowered base station (FAS) and the LIM are equipped with fluid antenna and liquid elements. LIM extends traditional RIS designs by incorporating reconfigurable spatial mobility into the metasurface structure \cite{lim1, lim2, lim3}, enabling each reflecting element to adjust its position in addition to its phase-shift control. This dual reconfigurability of spatial and electromagnetic domains offers unprecedented flexibility for adapting to dynamic wireless propagation paths. Note that as the inter-element spacing of an on–off-enabled metasurface decreases \cite{new2}, its behavior can be approximated to that of a quantized LIM. The proposed FAS-LIM system jointly leverages beam steering at BS, phase-shift at LIM, and spatial adjustment at both ends. Nonetheless, such integration introduces new challenges in system modeling and optimization due to the complex coupling between beamforming vectors, phase-shifts, and spatial positions. These interdependencies render the overall problem highly non-convex and difficult to solve if using the standard methods. To address this, we formulate a sum-rate maximization problem that jointly optimizes all control variables and develop a solution based on alternating optimization and other approximation techniques. The main contributions of this paper are summarized as follows:
\begin{enumerate}
    \item We have proposed a novel architecture in multi-user downlink systems that integrate fluid antenna arrays at BS and liquid elements at LIM, enabling joint optimization over transmit beamforming, phase-shifts configuration, and antenna/element position adjustment.
    
    \item  We aim for maximizing system sum-rate. We transform the non-convex problem into a tractable form using auxiliary variables, Lagrangian dual and quadratic transformation, successive convex approximation (SCA), and the penalty-based convex-concave procedure (PCCP) to tackle fractional forms, coupling variables, unit-modulus constraint, and spatial correlation functions. An alternating optimization technique is utilized to iteratively solve transmit beamforming, phase-shifts, and positions of FAS antennas and LIM elements. 
    
    \item Simulation results validate the effectiveness of the proposed solution, demonstrating that the FAS-LIM architecture achieves substantial rate performance gains over conventional designs, including fixed-antenna arrays, static RIS, and traditional BS-RIS architectures across various parameters.
\end{enumerate}

\begin{figure}[!t]
\centering
\includegraphics[width=2.5in]{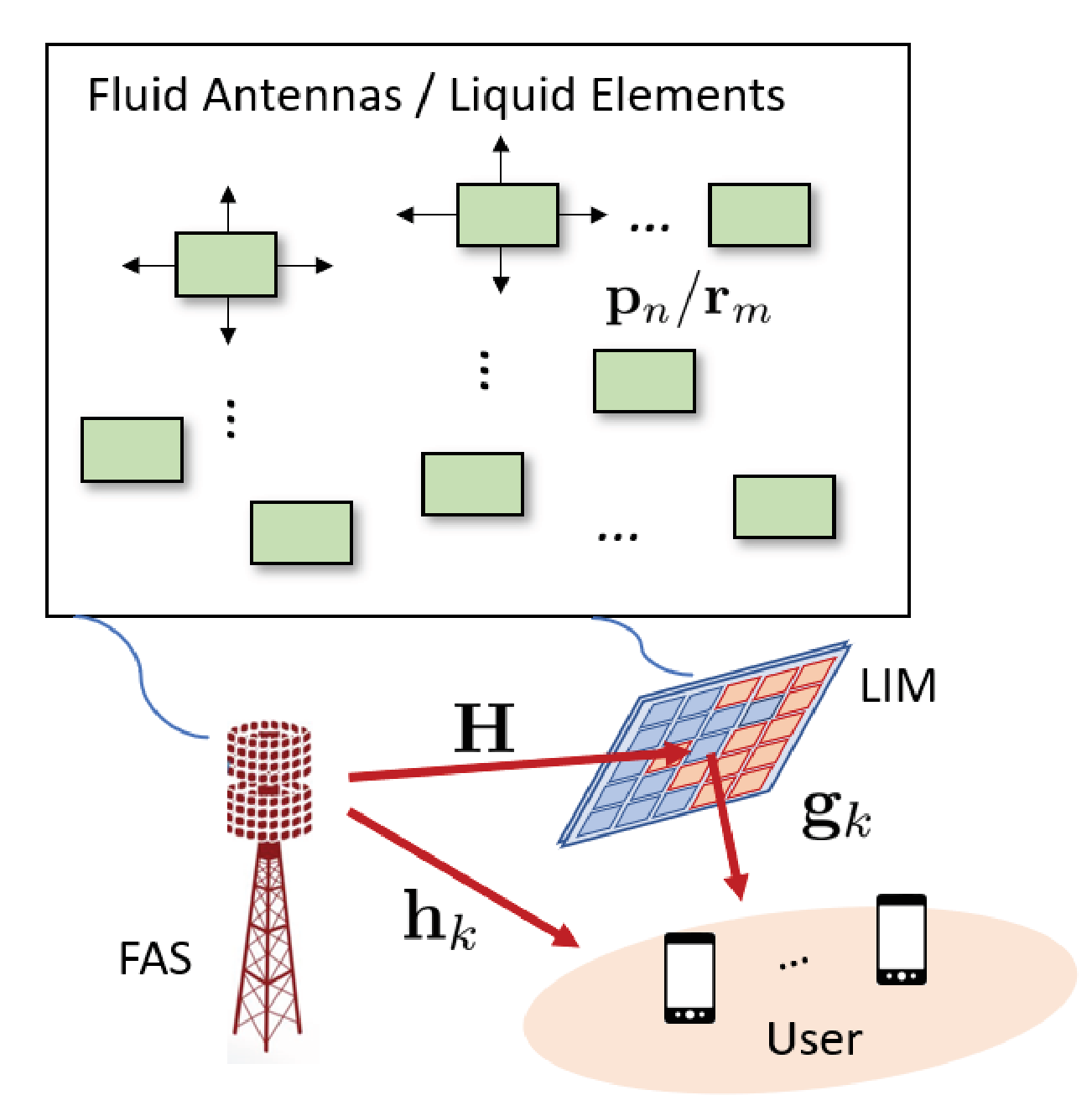}
\caption{The proposed architecture of LIM-assisted FAS.} \label{architecture}
\end{figure}

\section{System Model and Problem Formulation}

As shown in Fig. \ref{architecture}, we consider a downlink multi-user multiple-input single-output (MISO) system with one FAS and one LIM serving $K$ single-antenna users. The FAS is equipped with $N$ fluid antennas, each with a movable position $\mathbf{p}_n=(p_{x,n},p_{y,n}) \in \mathbb{R}^2$. The sizes of FAS and LIM are defined as $A_{\rm FA}=A_{{\rm FA},x}\cdot A_{{\rm FA},y}$ and $A_{\rm LM} = A_{{\rm LM},x} \cdot A_{{\rm LM},y}$, respectively. The LIM consists of $M$ liquid reflecting elements, associated with each element position of $\mathbf{r}_m =(r_{x,m},r_{y,m}) \in \mathbb{R}^2$ and its phase-shift $\theta_m$, where $|\theta_m| = 1$ and $\boldsymbol{\theta} = [\theta_1,\dots,\theta_M]^T$ is the vector of LIM phase-shifts. $T$ is the transpose operation. Note that each user $k \in \{1,\dots,K\}$ is located at position $\mathbf{u}_k$.

We consider the Rician fading channel model between the FAS and LIM as
\begin{align}
	\mathbf{H} = \sqrt{ \frac{h_0}{d_1^{\alpha}} }
\left( \sqrt{\frac{\kappa}{\kappa + 1}} \mathbf{H}_{\text{LoS}} + \sqrt{\frac{1}{\kappa + 1}} \mathbf{H}_{\text{NLoS}} \right),
\label{4}
\end{align}
where $h_0$ is the pathloss at the reference distance of 1 meter, $d_1$ is the distance between the FAS and LIM, and $\alpha$ is the pathloss exponent. Notation of $\kappa$ is the Rician factor adjusting the portion of LoS $\mathbf{H}_{\text{LoS}}$ and NLoS $\mathbf{H}_{\text{NLoS}}$. Following \eqref{4}, the channel between the FAS/LIM and user $k$ are defined as $\mathbf{h}_k = \sqrt{ \frac{h_0}{d_k^{\alpha}} }
\left( \sqrt{\frac{\kappa}{\kappa + 1}} \mathbf{h}_{k,\text{LoS}} + \sqrt{\frac{1}{\kappa + 1}} \mathbf{h}_{k,\text{NLoS}} \right)$ and $\mathbf{g}_k = \sqrt{ \frac{h_0}{d_{2,k}^{\alpha}} } \left( \sqrt{\frac{\kappa}{\kappa + 1}} \mathbf{g}_{k,\text{LoS}} + \sqrt{\frac{1}{\kappa + 1}} \mathbf{g}_{k,\text{NLoS}} \right)$  respectively, with $d_k$ and $d_{2,k}$ defined as the corresponding distances. While $\mathbf{h}_{k,\text{LoS}}$/$\mathbf{g}_{k,\text{LoS}}$ and $\mathbf{h}_{k,\text{NLoS}}$/$\mathbf{g}_{k,\text{NLoS}}$ stand for LoS and NLoS components of FAS/LIM to user $k$. The LoS components contain the steering vectors of FAS and those of LIM's incident and departure portions, respectively given by
\begin{align}
	[\mathbf{a}_{\rm FA}]_n &= e^{-j\frac{2\pi}{\lambda} (p_{x,n} \sin \varphi \cos \vartheta + p_{y,n} \sin \varphi \sin \vartheta)}, \\
	[\mathbf{a}_{{\rm LM},a}]_m &= e^{-j\frac{2\pi}{\lambda} (r_{x,m} \sin \varphi_a \cos \vartheta_a + r_{y,m} \sin \varphi_a \sin \vartheta_a)},
\end{align}
where $a\in\{t,r\}$ indicates the transmit or receiving steering vector of LIM, and $\lambda$ indicates the wavelength of the operating frequency. Notation of $\{\varphi, \vartheta\}$ indicates the azimuth and elevation angle-of-departure (AoD) of FAS, whereas $\{\varphi_a, \vartheta_a\}$ indicates the azimuth and elevation of AoD ($a=t$) and angle-of-arrival (AoA) ($a=r$) of LIM. Therefore, we can establish the LoS parts of FAS-LIM, and LIM-user, FAS-user respectively as
\begin{align}
\mathbf{H}_{\text{LoS}} = \mathbf{a}_{{\rm LM},r} \mathbf{a}_{\rm FA}^H, \ 
\mathbf{g}_{k,\text{LoS}} = \mathbf{a}_{{\rm LM},t}, \
\mathbf{h}_{k,\text{LoS}} = \mathbf{a}_{{\rm FA}},
\end{align}
where $H$ means hermitian operation.

Moreover, the spatial correlation between the FA and LIM elements are considered in the NLoS components, modeled by Jakes' model $[\mathbf{R}_q]_{i,j} = J_0\left( \frac{2\pi d_{i,j}}{\lambda} \right)$,
where $q\in\{ \O  , t, r\}$ stands for FAS's AoD and LIM's AoD/AoA, respectively. Notation $J_0(\cdot)$ is the Bessel function of the zero order \cite{fires} and $d_{i,j} = \lVert \mathbf{x}_i - \mathbf{x}_j \rVert, \forall \mathbf{x}\in\{\mathbf{p}, \mathbf{r}\}$ is the distance between the antenna/element $i$ and $j$ with $i\neq j$. The spatial correlation can be further written as $\mathbf{R}_q = \mathbf{U}_q \boldsymbol{\Lambda}_q \mathbf{U}_q^H$, where $\mathbf{U}_q$ is unitary matrix with eigenvectors and $ \boldsymbol{\Lambda}_q$ is the diagonal eigenvalue matrix. Consequently, the NLoS parts of FAS-LIM, FAS-user and LIM-user, are represented by $\mathbf{H}_{\text{NLoS}} = \mathbf{R}_{r}^{1/2} \bar{\mathbf{H}} \mathbf{R}^{1/2}$, 
$\mathbf{h}_{k,\text{NLoS}} = \mathbf{R}^{1/2} \bar{\mathbf{h}}_k$, and 
$\mathbf{g}_{k,\text{NLoS}} = \mathbf{R}_{t}^{1/2} \bar{\mathbf{g}}_k$ , respectively, where $\mathbf{R}_q^{1/2} = \mathbf{U}_q \boldsymbol{\Lambda}_q^{1/2} \mathbf{U}_q^H$. Note that $\bar{\mathbf{H}}$, $\bar{\mathbf{h}}_k$ and $\bar{\mathbf{g}}_k$ indicate the small-scale fading components, modeled as complex Gaussian distribution with
zero mean and unit variance. Then the combined channel between the FAS and user $k$ can be given by
\begin{align}
	\mathbf{h}_k^{\mathrm{eff}} = \mathbf{h}_{k}^H + \mathbf{g}_{k}^H \boldsymbol{\Theta} \mathbf{H},
\end{align}
where $\boldsymbol{\Theta} = \mathrm{diag}(\boldsymbol{\theta})$. The received signal of user $k$ is expressed as
\begin{align}
	y_k = \mathbf{h}_k^{\mathrm{eff}} \mathbf{w}_k x_k + \sum_{j \ne k} \mathbf{h}_k^{\mathrm{eff}} \mathbf{w}_j x_j + n_k,
\end{align}
where $\mathbf{w}_k \in \mathbb{C}^{N\times 1}$ is the beamforming vector for user $k$, $x_k \sim \mathcal{CN}(0,1)$ is the transmitted symbol, and $n_k \sim \mathcal{CN}(0,\sigma^2)$ is Gaussian noise. Therefore, the signal-to-interference-plus-noise ratio (SINR) of user $k$ can be given by
\begin{align}
	\gamma_k = \frac{|\mathbf{h}_k^{\mathrm{eff}} \mathbf{w}_k|^2}{\sum_{j \ne k} |\mathbf{h}_k^{\mathrm{eff}} \mathbf{w}_j|^2 + \sigma^2}.
\end{align}
Then the achievable rate\textsuperscript{\ref{note1}}\footnotetext[1]{The instant configuration is considered. Time overhead and power consumption of antenna and element movement are neglected here, which are potential factors to be considered in the future design.\label{note1}} is $R_k = \log_2(1 + \gamma_k)$. We aim for maximizing the total downlink rate by optimizing the FAS beamforming vector $\{\mathbf{w}_k\}$, LIM phase-shifts $\boldsymbol{\Theta}$, and FAS antenna/LIM element positions $\{\mathbf{p}_n\}$/$\{\mathbf{r}_m\}$, which is formulated as
\begingroup
\allowdisplaybreaks
\begin{subequations} \label{total_problem}
\begin{align}
& \max_{\substack{\{\mathbf{w}_k\}, \boldsymbol{\theta},  \\ \{\mathbf{p}_n\}, \{\mathbf{r}_m\}}} \quad \sum_{k=1}^K \log_2 \left( 1 + \gamma_k \right) \\
& \quad \text{s.t.}  \quad
  \sum_{k=1}^K \|\mathbf{w}_k\|^2 \le P_{\text{max}}, \label{con1} \\
	& \qquad\quad |\theta_m| = 1, \quad \forall m = 1, \dots, M, \label{con2} \\
	& \qquad\quad  \|\mathbf{p}_n - \mathbf{p}_{n'}\| \ge d_{\text{th}}^{\rm FA}, \quad \forall n \ne n', \label{con3} \\
	& \qquad\quad  \|\mathbf{r}_m - \mathbf{r}_{m'}\| \ge d_{\text{th}}^{\rm LM}, \quad \forall m \ne m', \label{con4} \\
	& \qquad\quad \mathbf{p}_n \in \mathcal{A}_p, \mathbf{r}_n \in \mathcal{A}_r, \forall n/m = 1, \dots, N/M. \label{con5}
\end{align}
\end{subequations}
\endgroup
In \eqref{con1}, maximum power is constrained by $P_{\text{max}}$. \eqref{con2} stands for LIM phase-shift constraint. Constraints \eqref{con3} and \eqref{con4} limit the minimum distance threshold between the antenna of FAS and elements of LIM as $d_{\text{th}}^{\rm FA}$ and $d_{\text{th}}^{\rm LM}$, respectively. \eqref{con5} limits the positions within the antenna array $\mathcal{A}_p$ and metasurface $\mathcal{A}_r$. The problem is non-linear and non-convex, which is challenging. Therefore, we employ alternating optimization to iteratively solve the subproblems, elaborated as follows.

\section{Proposed Solution}

\subsection{Problem Transformation}

We reformulate the problem using the Lagrangian dual reformulation \cite{dstar} with the auxiliary parameter $
\boldsymbol{\gamma}=\{\bar{\gamma}_k | \forall k\in\{1,...,K\} \}$, derived by
\begin{align} \label{fr}
	f_{r}(\boldsymbol{\Xi}, \boldsymbol{\gamma}) = \sum_{k=1}^{K} \log(1+\bar{\gamma}_k) - 
\sum_{k=1}^{K} \bar{\gamma}_k 
	+ \sum_{k=1}^{K} \frac{(1+\bar{\gamma}_k) | \mathbf{h}_k^{\mathrm{eff}} \mathbf{w}_k|^2 }{ \sum_{j=1}^{K} |\mathbf{h}_k^{\mathrm{eff}} \mathbf{w}_j|^2 + \sigma^2 },
\end{align}
where $\boldsymbol{\Xi} = \{\mathbf{w}_k, \boldsymbol{\theta}, \mathbf{p}_n, \mathbf{p}_m\}$ indicates the parameter set. Notation $\bar{\gamma}_k$ can be derived by setting $\frac{\partial f_{r}(\mathbf{w},  \boldsymbol{\theta})}{ \partial \bar{\gamma}_k}=0$, resulting in $\bar{\gamma}_k^{\star} = \frac{| \mathbf{h}_k^{\mathrm{eff}} \mathbf{w}_k|^2}{\sum_{j=1, j\neq k}^{K} | \mathbf{h}_k^{\mathrm{eff}} \mathbf{w}_j|^2 + \sigma^2}$. Note that the optimal $\bar{\gamma}_k$ is the SINR of user $k$, where this term can be updated by using the solution at previous iteration. When $\bar{\gamma}_k$ is fixed, only the last term of \eqref{fr} has a sum-of-ratio form. Using the quadratic transform \cite{new1} yields
\begingroup
\allowdisplaybreaks
\begin{align} \label{y_lagrang}
	f_{r}(\boldsymbol{\Xi}, \boldsymbol{\gamma}, \mathbf{y}) &= \sum_{k=1}^{K} 2 \bar{y}_k \sqrt{ (1+ \bar{\gamma}_k) | \mathbf{h}_k^{\mathrm{eff}} \mathbf{w}_k|^2} \notag\\
	& - \underbrace{\sum_{k=1}^{K} \bar{y}_k^2 \left( \sum_{j=1 }^{K} |\mathbf{h}_k^{\mathrm{eff}} \mathbf{w}_j |^2 +\sigma^2 \right)}_{\triangleq \bar{f}_r (\boldsymbol{\Xi}, \mathbf{y})} + c_{\gamma},
\end{align}
\endgroup
where $\mathbf{y}=\{\bar{y}_k| \forall k \in \{1,...,K \} \}$ and $c_{\gamma} = \sum_{k=1}^{K} \log(1+\bar{\gamma}_k) - 
\sum_{k=1}^{K} \bar{\gamma}_k$ is the constant term related to $\bar{\gamma}_k$, where the closed-form of $\bar{y}_k$ is derived following the same process as that of $\bar{\gamma}_k$, i.e., $\bar{y}_k^{\star} = \frac{\sqrt{ (1+\bar{\gamma}_k) | \mathbf{h}_k^{\mathrm{eff}} \mathbf{w}_k|^2 }}{\sum_{j \in \mathcal{K}, j\neq k} |\mathbf{h}_k^{\mathrm{eff}} \mathbf{w}_j|^2 + \sigma^2}$. Then the transformed problem becomes related to $\boldsymbol{\Xi}$. We adopt block coordinate descent (BCD) method to solve the respective subproblems as follows.

\subsection{Subproblem for Solving Beamforming $\mathbf{w}_{k}$}

Given fixed LIM phase-shifts $\boldsymbol{\theta}$,  FAS/LIM antenna/element positions $\{\mathbf{p}_n, \mathbf{r}_m\}$, we aim for optimizing the FAS beamforming vectors $\{\mathbf{w}_k\}$ to maximize the sum-rate. The corresponding optimization problem is
\begin{align} \label{sub1}
\max_{\{\mathbf{w}_k\}} \quad & f_{r}(\boldsymbol{\Xi}, \boldsymbol{\gamma}, \mathbf{y}) \quad
\text{  s.t.} \ \eqref{con1}.
\end{align}
This problem is non-convex due to the first term of \eqref{y_lagrang}. To make it more tractable, we convexify the term of $|\mathbf{h}_k^{\mathrm{eff}} \mathbf{w}_k|$ by using SCA as
\begin{align} \label{sub1_c3}
	g_w(\mathbf{w}_k) \triangleq |\mathbf{h}_{k}^{\rm eff} \mathbf{w}_{k}^{(t)}| + \frac{1}{|\mathbf{h}_{k}^{\rm eff} \mathbf{w}_{k}^{(t)}|} \Re \left\{ (\mathbf{w}_k^{(t)})^H \mathbf{H}_k \left( \mathbf{w}_k \!-\! \mathbf{w}_k^{(t)} \right) \right\},
\end{align}
where $\mathbf{H}_k = (\mathbf{h}_k^{\rm eff})^H \mathbf{h}_k^{\rm eff}$ and $\Re \{ \cdot \}$ indicates real part of a variable. Note that the second term in \eqref{y_lagrang} is concave w.r.t. $\mathbf{w}_k$ and needs no approximation. Then the problem now becomes
\begin{align} \label{sub1_2}
\max_{\substack{\{\mathbf{w}_k\}}} \quad & \sum_{k=1}^{K} 2 \bar{y}_k \sqrt{ 1+ \bar{\gamma}_k} \cdot g_w(\mathbf{w}_k) - \bar{f}_r (\boldsymbol{\Xi}, \mathbf{y}) \
\text{  s.t. } \ \eqref{con1},
\end{align}
which is convex and solvable via standard optimization tools.

\subsection{Subproblem for Solving LIM Phase-Shift $\boldsymbol{\theta}$}

Given fixed parameters of $\{ \mathbf{w}_k, \mathbf{p}_n, \mathbf{r}_m\}$, we proceed to optimize the LIM phase-shifts $\boldsymbol{\theta}$. First, we define $\mathbf{h}_k^{\mathrm{eff}}(\boldsymbol{\theta}) = \mathbf{h}_{k}^H + \boldsymbol{\theta}^T \mathbf{D}_k$ and $\mathbf{D}_k \triangleq \mathrm{diag}(\mathbf{g}_{k}) \mathbf{H}$. Let us define $s_{k,j} = \mathbf{h}_{k}^H \mathbf{w}_j$ and $\mathbf{v}_{k,j} = \mathbf{D}_k \mathbf{w}_j$. Then we have
$f_{k,j}(\boldsymbol{\theta}) = | \mathbf{h}_k^{\mathrm{eff}}(\boldsymbol{\theta}) \mathbf{w}_j | = | s_{k,j} + \boldsymbol{\theta}^T \mathbf{v}_{k,j} |$. At iteration $t$, define $\mu_{k,j}^{(t)} = s_{k,j} + (\boldsymbol{\theta}^{(t)})^T \mathbf{v}_{k,j}$. Then the SCA approximation of $f_{k,j}(\boldsymbol{\theta})$ can be given by
\begin{align} \label{conn_t}
	\tilde{f}_{k,j}(\boldsymbol{\theta}) = |\mu_{k,j}^{(t)}| + \frac{1}{|\mu_{k,j}^{(t)}|} \Re \left\{ \mu_{k,j}^{(t)\ast} \mathbf{v}_{k,j}^T (\boldsymbol{\theta} - \boldsymbol{\theta}^{(t)}) \right\}.
\end{align}
Substituting \eqref{conn_t} by setting $j=k$ into the first non-convex term of \eqref{y_lagrang} yields the convex objective. Additionally, we utilize PCCP to handle the non-convex unit-modulus constraint $|\theta_m| = 1$ in \eqref{con2} by equivalently relaxing it with a pair of constraints $|\theta_m|^2 \le 1 + c_m$ and $|\theta_m|^2 \ge 1 - c_m$, where $c_m$ is the penalty factor of PCCP \cite{dstar}. Then the non-convex inequality $|\theta_m|^2 \ge 1 - c_m$ is therefore approximated by SCA as
\begin{align} \label{sub2_1}
	|\theta_m^{(t)}|^2 + 2 \Re \left\{ \theta_m^{(t)\ast} (\theta_m-\theta_m^{(t)}) \right\} \ge 1 - c_m.
\end{align}
Then the problem optimizing LIM phase-shifts is formulated as
\begingroup
\allowdisplaybreaks
\begin{subequations} \label{sub2}
\begin{align}
& \max_{\substack{\boldsymbol{\theta},c_m}} \ \sum_{k=1}^{K} 2 \bar{y}_k \sqrt{ 1+ \bar{\gamma}_k} \cdot \tilde{f}_{k,k}(\boldsymbol{\theta}) - \bar{f}_r (\boldsymbol{\Xi}, \mathbf{y}) - \xi \sum_{m=1}^M c_m \\
& \quad \text{s.t.}  \quad \eqref{sub2_1}, \qquad |\theta_m|^2 \le 1 + c_m, \quad \forall m = 1, \dots, M, \\
	& \qquad\quad c_m \ge 0, \quad \forall m = 1, \dots, M.
\end{align}
\end{subequations}
\endgroup
Note that the penalty weight $\xi > 0$ controls how strongly we enforce the unit-modulus constraint. The problem \eqref{sub2} is convex and can be solved via arbitrary optimization tools.

\subsection{Subproblem for Joint FAS-LIM Positions $\{\mathbf{p}_n, \mathbf{r}_m \}$ }

Given fixed parameters of $\{\mathbf{w}_k, \boldsymbol{\theta}\}$, we optimize the FAS-FIM positions of $\{\mathbf{p}_n , \mathbf{r}_m \}$. We can observe that it is challenging to solve the highly-coupled and non-linear terms in both exponentials and Bessel functions. We rewrite the effective channel as
\begin{align*}
	\mathbf{h}_k^{\mathrm{eff}}(\mathbf{p}, \mathbf{r}) = \mathbf{h}_{k}^H (\mathbf{p}) + \mathbf{g}_{k}^H(\mathbf{r}) \cdot \boldsymbol{\Theta} \cdot \mathbf{H}(\mathbf{p}, \mathbf{r}).
\end{align*}
We define $g_{k,j,\nu}(\mathbf{p}, \mathbf{r}) = |\mathbf{h}_k^{\mathrm{eff}}(\mathbf{p}, \mathbf{r}) \mathbf{w}_j|^{\nu},\forall \nu\in\{1,2\}$, with its SCA approximation at stationary point $(\mathbf{p}^{(t)}, \mathbf{r}^{(t)})$ derived as
\begingroup
\allowdisplaybreaks
\begin{align} \label{grad_g}
\tilde{g}_{k,j,\nu}(\mathbf{p}, \mathbf{r}) &= g_{k,j,\nu}^{(t)}(\mathbf{p}, \mathbf{r})  + \mathfrak{R}\left\lbrace \left( \nabla_{\mathbf{p}_n} g_{k,j,\nu}^{(t)} \right) ^H (\mathbf{p}_n - \mathbf{p}_n^{(t)}) \right\rbrace \notag \\ 
	& \qquad + \mathfrak{R}\left\lbrace \left( \nabla_{\mathbf{r}_m} g_{k,j,\nu}^{(t)} \right)^H (\mathbf{r}_m \!-\! \mathbf{r}_m^{(t)} ) \right\rbrace ,
\end{align}
\endgroup
where the derivation of gradients $\nabla_{\mathbf{p}_n} g_{k,j,\nu}^{(t)}$ and $\nabla_{\mathbf{r}_m} g_{k,j,\nu}^{(t)}$ can be found in Appendix. Additionally, inter-antenna and inter-element spacing constraints in \eqref{con3} and \eqref{con4} are respectively approximated by SCA as
\begingroup
\allowdisplaybreaks
\begin{align}
	& \|\mathbf{p}_n - \mathbf{p}_{n'}\| \ge \|\Delta_{nn'}^{(t)}\| + (\Delta_{nn'}^{(t)})^T /  \|\Delta_{nn'}^{(t)}\| \cdot \notag \\
	&\qquad\qquad\qquad\qquad \left[ (\mathbf{p}_n - \mathbf{p}_n^{(t)}) - (\mathbf{p}_{n'} - \mathbf{p}_{n'}^{(t)}) \right],	 \label{sub3_1} \\
	& \|\mathbf{r}_m - \mathbf{r}_{m'}\| \ge \|\bar{\Delta}_{mm'}^{(t)}\| +  (\bar{\Delta}_{mm'}^{(t)})^T / \|\bar{\Delta}_{mm'}^{(t)}\| \cdot \notag \\
	&\qquad\qquad\qquad\qquad \left[ (\mathbf{r}_m - \mathbf{r}_m^{(t)}) - (\mathbf{r}_{m'} - \mathbf{r}_{m'}^{(t)}) \right].  \label{sub3_2}
\end{align}
\endgroup
where $\Delta_{nn'}^{(t)} \triangleq \mathbf{p}_n^{(t)} - \mathbf{p}_{n'}^{(t)}$ and $\bar{\Delta}_{mm'}^{(t)} \triangleq \mathbf{r}_m^{(t)} - \mathbf{r}_{m'}^{(t)}$. Then the optimization problem becomes
\begingroup
\allowdisplaybreaks
\begin{subequations} \label{sub3}
\begin{align}
	& \max_{\substack{\{\mathbf{p}_n\},\, \{\mathbf{r}_m\}}} \quad 
	\sum_{k=1}^{K} 2 \bar{y}_k \sqrt{ 1+ \bar{\gamma}_k} \cdot \tilde{g}_{k,k,1}(\mathbf{p}, \mathbf{r}) \notag \\
	&\qquad\qquad\qquad\qquad - \sum_{k=1}^{K} \bar{y}_k^2 \left( \sum_{j=1 }^{K} \tilde{g}_{k,j,2}(\mathbf{p}, \mathbf{r}) +\sigma^2 \right)	
	 \\
& \text{s.t.} 
	\quad \eqref{con5}, \eqref{sub3_1}, \eqref{sub3_2},
\end{align}
\end{subequations}
\endgroup
which is joint convex with respect to $\{\mathbf{p}, \mathbf{r}\}$ and can be solved by using any convex optimization tools.

\begin{algorithm}[!t]
\caption{Proposed Solution for FAS-LIM}
\footnotesize
\SetAlgoLined
\DontPrintSemicolon
\label{alg}
\begin{algorithmic}[1]
\STATE Randomly initialize all parameters $\{\mathbf{w}_{k}, \boldsymbol{\theta}, \mathbf{p}_{n},\mathbf{r}_{m}, c_m\}$
\REPEAT
	\STATE Update $\bar{\gamma}_k$ and $\bar{y}_k$ using previous solutions $\{\mathbf{w}_{k}^{(t)}, \boldsymbol{\theta}^{(t)}, \mathbf{p}_{n}^{(t)},\mathbf{r}_{m}^{(t)}\}$
	\STATE Solve FAS beamforming $\mathbf{w}_{k}$ in problem \eqref{sub1_2}
	\STATE Solve LIM configuration $\boldsymbol{\theta}$ in problem \eqref{sub2}
	\STATE Solve antenna/element positions $\{\mathbf{p}_{n},\mathbf{r}_{m} \}$ in problem \eqref{sub3}
\UNTIL Convergence of total solution
\STATE {\bf return} $\{\mathbf{w}_{k}, \boldsymbol{\theta}, \mathbf{p}_{n},\mathbf{r}_{m}\}$.
\end{algorithmic}
\end{algorithm}

\begin{figure*}[!ht]
	\centering
	\subfigure[]{\includegraphics[width=1.7 in]{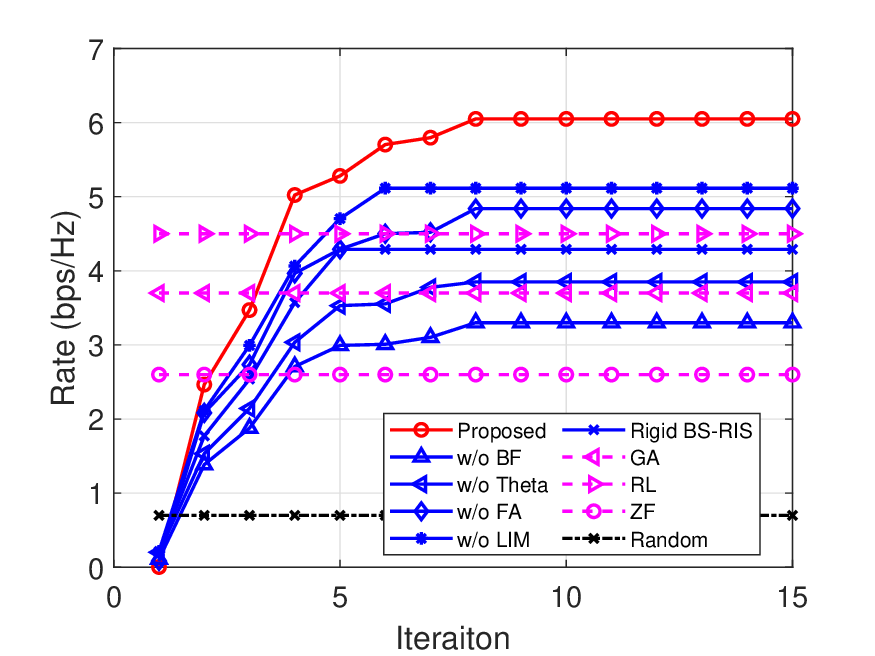} \label{fig1}}
	\subfigure[]{\includegraphics[width=1.7 in]{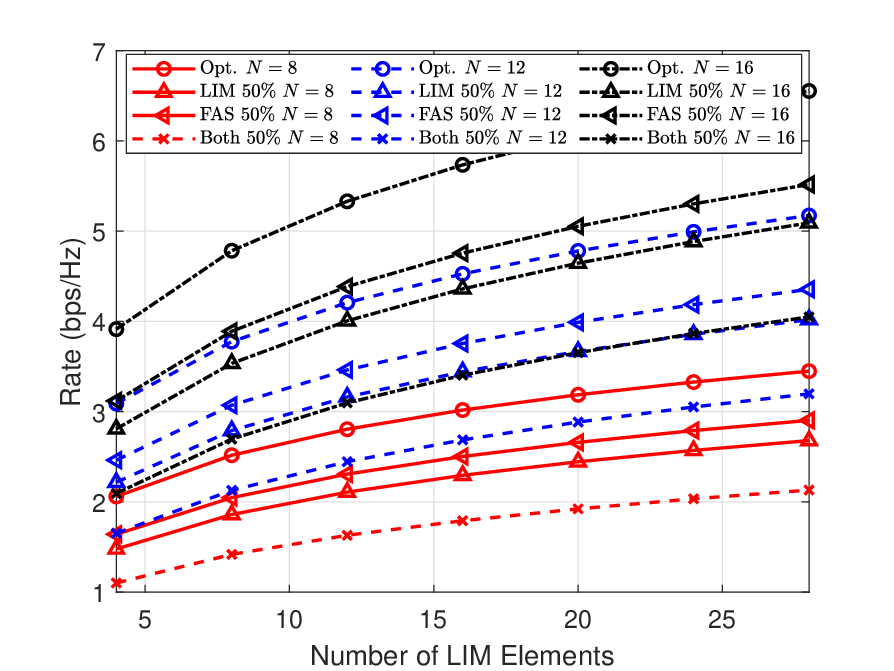} \label{fig2}}
	\subfigure[]{\includegraphics[width=1.7 in]{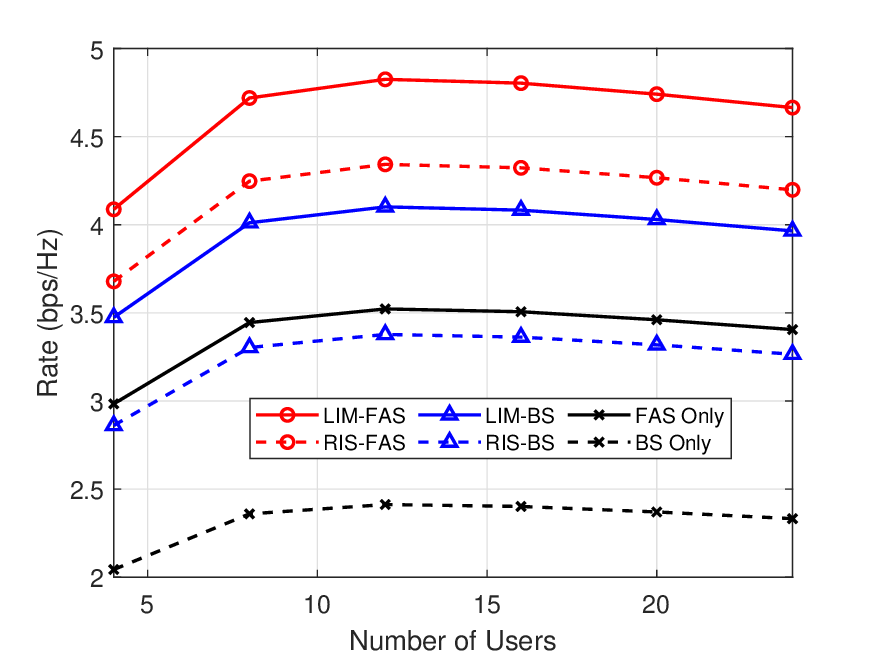} \label{fig3}}
	\subfigure[]{\includegraphics[width=1.7 in]{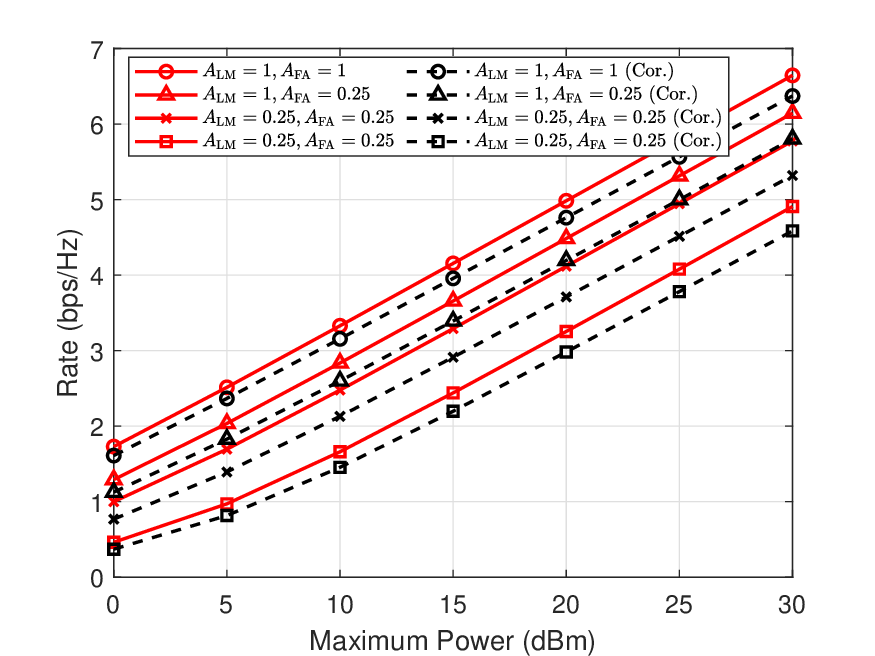} \label{fig4}}	
	\caption{(a) Convergence. (b) Rate with various numbers of antennas/elements. (c) Rate with various numbers of users. (d) Rate with different maximum power.}
\end{figure*}

\subsection{Computational Complexity Analysis}

We analyze the computational complexity of the proposed alternating optimization framework, where each subproblem is solved using an interior-point method. The complexity per subproblem depends on the number of variables, equality/inequality constraints, and the log-barrier terms in the interior-point formulation. Subproblem 1 in \eqref{sub1} optimizes $N$-dimensional complex beamformers $\{\mathbf{w}_k\}_{k=1}^K$ along with auxiliary variables $\{\bar{y}_k, \bar{\gamma}_k\}$. Note that auxiliary variables are updated directly using the closed-forms. The total number of real variables is $2KN$. Using interior-point method to solve the convex problem, the worst-case complexity is in an order of
$
\mathcal{O}\left( \left( 2KN\right)^3 \cdot \log\left( \frac{1}{\epsilon} \right) + 2K \right),
$
where $\epsilon$ is the accuracy of convergence. Subproblem 2 in \eqref{sub2} optimizes $\boldsymbol{\theta} \in \mathbb{C}^M$ with $M$ real variables, and slack variable $\{c_m\}_{m=1}^M$. The total number of real variables is $2M$. Then its computational complexity is obtained as
$
\mathcal{O}\left( \sqrt{M} \cdot \left( 2M \right)^3 \cdot \log\left( \frac{1}{\epsilon} \right) \right).
$
Subproblem 3 in \eqref{sub3} has total variables of $3(N + M)$. Therefore, its computational complexity is acquired as
$
\mathcal{O} \left( \sqrt{N^2 + M^2} \cdot \left( 3(N + M)\right)^3 \cdot \log\left( \frac{1}{\epsilon} \right) \right).
$
Let $I_{\text{outer}}$ be the number of outer SCA iterations. Then the total complexity can be denoted as
$ \mathcal{O}\left( I_{\text{outer}} \cdot \left( C_{\mathbf{w}} + C_{\boldsymbol{\theta}} + C_{\mathbf{p},\mathbf{r}} \right) \right),
$ where $C_{\mathbf{w}}$, $C_{\boldsymbol{\theta}}$, and $C_{\mathbf{p},\mathbf{r}}$ denote the per-iteration complexity of each subproblem as derived above, respectively.

\section{Simulation Results}

Simulation results are provided to validate the effectiveness of FAS-LIM architecture and proposed solution. The FAS and LIM are located at $(0,0)$ and $(50, 20)$ m, respectively. Users are uniformly and randomly distributed within a radius of $10$ m centered at the location $(100, 0)$ m. The number of antennas/elements/users are set to $N=16$, $M=16$, and $K=8$, respectively. Other related parameters are set as follows: $h_0 = -20$ dB, $\alpha = 2.2$, $\kappa=3$ , $\lambda = 0.1$ m, $\sigma^2 = -95$ dBm, $P_{\text{max}}=30$ dBm, $A_{\rm LM/FA} = 1$ m$^2$, $d_{\text{th}}^{\rm FA/LM} = 0.1$ m, $\xi=10^3$, $I_{\text{outer}}=15$.

Fig.~\ref{fig1} illustrates the rate convergence behavior of the proposed algorithm. It can be observed that the proposed scheme converges to the maximum sum-rate at around $9$-th iterations. Note that "w/o" indicates random parameter generation. In contrast, the baselines of random beamforming (w/o BF), random phase-shift (w/o Theta), random positions of fluid antennas (w/o FA), the absence of LIM (w/o LIM), and rigid arrays at BS-RIS, all exhibit noticeable rate degradation. Conventional optimization baselines such as zero-forcing (ZF) \cite{BM_ZF}, heuristic genetic algorithm (GA) \cite{GA}, and quantized reinforcement learning (RL) \cite{my} fail to cope with the high-dimensional and non-convex joint optimization, resulting in a lower rate than that of the proposed solution for FAS-LIM. The random baseline consistently maintains the lowest rate, which demonstrates the critical role of proper joint optimization design.

Fig.~\ref{fig2} investigates the achievable rate under varying numbers of LIM elements and FAS antennas with different levels of partially-configurable antennas or elements. The fully optimized scheme (Opt.) outperforms partially cases confirming the benefit of joint optimization. Even when only 50\% of either the FAS antennas or LIM elements are adjustable, significant performance gains can still be observed. However, the case where only 50\% of both FAS antennas and LIM elements are adjustable experiences noticeable rate degradation, highlighting the necessity of full configurability on at least one side for optimal performance. Furthermore, increasing the number of antennas and elements enhances the spatial degrees of freedom for position adjustment, thereby contributing to improved rate performance.

Fig.~\ref{fig3} presents the system sum-rate with different numbers of users. It is evident that the proposed FAS-LIM architecture achieves the highest rate across all numbers of users, outperforming the other hybrid combinations of LIM-BS, RIS-FAS, and RIS-BS. Systems with only FAS or BS arrays exhibit much lower rates due to the lack of intelligent reconfigurability. Moreover, the performance trend exhibits saturation as the number of users increases, indicating a performance limit imposed by the confined power budget and surface sizes. Further increasing the excessive number of users will lead to rate degradation due to insufficient resources.

Fig.~\ref{fig4} analyzes the impact of maximum transmit power on the achievable rate under different LIM/FAS surface sizes and the case with/without spatial correlation factors. As expected, increasing power leads to a nearly linear improvement in rate. Also, larger LIM and FAS surface sizes offer higher rate performance, benefited from its higher flexibility of adjusting positions. However, when considering NLoS spatial correlation (Cor.), the system suffers from the reduced rate of around 5\% loss. These results emphasize the importance of accounting for the spatial correlation effects in practical deployments and highlight the benefit of larger sizes of FAS-LIM surfaces.

\section{Conclusion}

In this paper, we have introduced a novel FAS-LIM-assisted MISO downlink system, where both the BS and the LIM are respectively equipped with fluid antennas and liquid elements capable of performing dynamic spatial repositioning. By jointly optimizing the FAS beamforming, LIM phase-shifts, and locations of fluid antenna and liquid elements, the proposed architecture achieves the enhanced electromagnetic and spatial reconfigurability. To tackle the resulting non-convex optimization problem, we employ an alternating optimization algorithm with auxiliary variable reformulation, Lagrangian dual and quadratic transformation, SCA and PCCP methods. Simulation results have confirmed that the proposed FAS-LIM framework yields substantial sum-rate improvements over benchmarks with static RIS or fixed antenna arrays as well as other baseline methods. Such findings underscore the promise of combining FAS with LIM for intelligent and adaptive wireless networks.

\appendix

We breakdown the effective channel $\mathbf{h}^{\rm eff}_{k}$ as
\begin{align} \label{gra}
	& \mathbf{h}^{\rm eff}_{k}(\mathbf{p}, \mathbf{r}) = 
	c_1 \mathbf{h}_{k,\text{LoS}} + c_2 \mathbf{g}^{H}_{k, \text{LoS}} \boldsymbol{\Theta}\mathbf{H}_{\text{LoS}}  
	+ c_3 \mathbf{h}_{k,\text{NLoS}} 
	+ \notag\\ & 
	c_4 \mathbf{g}^{H}_{k, \text{NLoS}} \boldsymbol{\Theta}\mathbf{H}_{\text{NLoS}}
	+
	c_5 \mathbf{g}^{H}_{k, \text{LoS}} \boldsymbol{\Theta}\mathbf{H}_{\text{NLoS}}+
	c_5 \mathbf{g}^{H}_{k, \text{NLoS}} \boldsymbol{\Theta}\mathbf{H}_{\text{LoS}},
\end{align}
where 
	$c_1 = \sqrt{\frac{h_0 \kappa}{d_k^{\alpha}(\kappa+1)}}$, 
	$c_2 = \sqrt{\frac{h_0^2 \kappa^2}{(d_1 d_{2,k})^{\alpha} (\kappa+1)^2}}$,
	$c_3 = \sqrt{\frac{h_0}{d_k^{\alpha}(\kappa+1)}}$, 
	$c_4 = \sqrt{\frac{h_0^2}{(d_1 d_{2,k})^{\alpha} (\kappa+1)^2}}$, and
	$c_5 = \sqrt{\frac{h_0^2 \kappa}{(d_1 d_{2,k})^{\alpha} (\kappa+1)^2}}$ are constants. Define $g_{k,j,\nu}(\mathbf{p}, \mathbf{r}) = |s_{k,j}|^{\nu}$ where $\nu\in\{1,2\}$ and $s_{k,j} = \mathbf{h}_k^{\mathrm{eff}}(\mathbf{p}, \mathbf{r}) \mathbf{w}_j $. Using the chain rule, we then have $\nabla_{\mathbf{x}} g_{k,j,\nu} = \frac{1}{|s_k,j|} \mathfrak{R}\{ ( \nabla_{\mathbf{x}} \mathbf{h}^{\rm eff}_{k} \mathbf{w}_j)^* s_{k,j}\}$ if $\nu=1$ and $\nabla_{\mathbf{x}} g_{k,j,\nu} = 2 \mathfrak{R}\{ ( \nabla_{\mathbf{x}} \mathbf{h}^{\rm eff}_{k} \mathbf{w}_j)^* s_{k,j}\}$ if $\nu=2$, where $\mathbf{x} \in \{ \mathbf{p}_n, \mathbf{r}_m \}$. We will derive the gradient of $\mathbf{h}^{\rm eff}_{k}$ in terms of $\mathbf{p}_n$ and $\mathbf{r}_m$ as follows.

\subsubsection{Gradient to $\mathbf{p}_n$}

We can know from \eqref{gra} that $\mathbf{h}_{k,\text{LoS/NLoS}}$ and $\mathbf{H}_{\text{LoS/NLoS}}$ are related to $\mathbf{p}_n$. As for the LoS part, we have 
\begin{align} \label{gr1}
	\nabla_{\mathbf{p}_n} \mathbf{h}_{k,\text{LoS}} =  \nabla_{\mathbf{p}_n} \mathbf{a}_{\rm FA} = -j \frac{2\pi}{\lambda} \mathbf{u}_{\rm FA} \cdot [\mathbf{a}_{\rm FA}]_n, 
\end{align}
where $\mathbf{u}_{\rm FA} = \left[\begin{matrix}
\sin \varphi \cos \vartheta \\
\sin \varphi \sin \vartheta
\end{matrix}\right]$. Similarly, we have $\nabla_{\mathbf{p}_n} \mathbf{H}_{\text{LoS}} = \mathbf{a}_{{\rm LM},r} \cdot \nabla^H_{\mathbf{p}_n} \mathbf{a}_{\rm FA} $. We now proceed to solve NLoS part. However, it leads to a difficulty of solving square-root matrix with Bessel functions, i.e., $\nabla_{\mathbf{p}_n} \mathbf{h}_{k,\text{NLoS}} = ( \nabla_{\mathbf{p}_n} \mathbf{R}^{1/2}) \bar{\mathbf{h}}_k $. Employing Sylvester equation \cite{syl} to compute $\nabla_{\mathbf{p}_n} \mathbf{R}^{1/2}$ yields
\begin{align} \label{gr2}
	 \mathbf{R}^{1/2} \cdot \mathbf{X} + \mathbf{X} \cdot \mathbf{R}^{1/2} = \nabla_{\mathbf{p}_n} \mathbf{R} \Leftrightarrow \mathbf{X} = \nabla_{\mathbf{p}_n} \mathbf{R}^{1/2},
\end{align}
where the gradient of $\nabla_{\mathbf{p}_n} \mathbf{R}$ is obtained as 
\begin{align} \label{gr3}
	\nabla_{\mathbf{p}_n} [\mathbf{R}]_{i,j} =
\left\{\begin{array}{ll}
	-\frac{2 \pi}{\lambda} J_1 (\frac{2\pi}{\lambda} \lVert \mathbf{p}_n- \mathbf{p}_j \rVert) \cdot \frac{\mathbf{p}_n- \mathbf{p}_j}{ \lVert \mathbf{p}_n- \mathbf{p}_j \rVert },  & \mbox{if } i=n, \\
	-\frac{2 \pi}{\lambda} J_1 (\frac{2\pi}{\lambda} \lVert \mathbf{p}_i- \mathbf{p}_n \rVert) \cdot \frac{\mathbf{p}_i- \mathbf{p}_n}{ \lVert \mathbf{p}_n - \mathbf{p}_i \rVert },  & \mbox{if } j=n,
\end{array} \right.
\end{align}	
where $J_1(\cdot)$ is the Bessel function of the first order \cite{fires}. The closed-form of $\mathbf{X}$ is based on Kronecker product vectorization, given by
\begin{align} \label{gr4}
	{\rm vec}(\mathbf{X}) = (\mathbf{I} \otimes \mathbf{R}^{1/2} + (\mathbf{R}^{1/2})^T \otimes \mathbf{I})^{-1} \cdot {\rm vec}( \nabla_{\mathbf{p}_n} \mathbf{R} ), 
\end{align}
where $ \mathbf{X} ={\rm unvec}( {\rm vec}(\mathbf{X}) )$. ${\rm vec}(\cdot)$ and ${\rm unvec}(\cdot)$ vectorizes and unvectorizes the matrix, respectively. Similarly, we have $\nabla_{\mathbf{p}_n} \mathbf{H}_{\text{NLoS}} = \mathbf{R}_{r}^{1/2} \bar{\mathbf{H}} \cdot  \nabla_{\mathbf{p}_n} \mathbf{R}^{1/2} $. Combining above gradients yields the final total gradient $\nabla_{\mathbf{p}_n} \mathbf{h}^{\rm eff}_{k}$ and the corresponding $\nabla_{\mathbf{p}_n} g_{k,j,\nu}$.

\subsubsection{Gradient to $\mathbf{r}_m$}
Following the same derivation method in \eqref{gr1}, we can obtain $\nabla_{\mathbf{r}_m} \mathbf{g}_{k,\text{LoS}} = \nabla_{\mathbf{r}_m} \mathbf{a}_{{\rm LM},r} =  -j \frac{2\pi}{\lambda} \mathbf{u}_{{\rm LM},a} \cdot [\mathbf{a}_{{\rm LM},a}]_m, $, where $\mathbf{u}_{{\rm LM},a} = \left[\begin{matrix}
\sin \varphi_a \cos \vartheta_a \\
\sin \varphi_a \sin \vartheta_a
\end{matrix}\right], \forall a\in\{t,r\}$. Similarly, we have $\nabla_{\mathbf{r}_m} \mathbf{H}_{k,\text{LoS}} =  \nabla_{\mathbf{r}_m} \mathbf{a}_{{\rm LM},r} \cdot \mathbf{a}_{\rm FA}^H$. The gradients of NLoS parts can be obtained as $\nabla_{\mathbf{r}_m} \mathbf{g}_{k,\text{NLoS}} =  (\nabla_{\mathbf{r}_m} \mathbf{R}_t^{1/2}) \bar{\mathbf{g}}_k$ and $\nabla_{\mathbf{r}_m} \mathbf{H}_{\text{NLoS}} =  (\nabla_{\mathbf{r}_m} \mathbf{R}_r^{1/2}) \bar{\mathbf{H}} \mathbf{R}^{1/2}$, where $\nabla_{\mathbf{r}_m} \mathbf{R}_{a}^{1/2}, \forall a\in\{t,r\}$ follows the same process in \eqref{gr3}. Since \eqref{gra} possesses the coupled terms, the product rule for derivatives is utilized, i.e., $\nabla (c_i \mathbf{A}^H \boldsymbol{\Theta} \mathbf{B}) = c_i \cdot [ (\nabla\mathbf{A})^H \cdot \boldsymbol{\Theta} \mathbf{B} + \mathbf{A}^H \boldsymbol{\Theta}\cdot  (\nabla\mathbf{B}) ]$, where $\mathbf{A}= \mathbf{g}_{k,\text{LoS/NLoS}}$ and $\mathbf{B} = \mathbf{H}_{\text{LoS/NLoS}}$. Combining gradients above yields the final gradient $	\nabla_{\mathbf{r}_m} \mathbf{h}^{\rm eff}_{k}$ and the corresponding $	\nabla_{\mathbf{r}_m} g_{k,j,\nu}$. 

This completes derivations of total gradients $\nabla_{\mathbf{x}} g_{k,j,\nu},\forall\mathbf{x}\in\{ \mathbf{p}_n, \mathbf{r}_m \}$ by substituting $\nabla_{\mathbf{x}} g_{k,j,\nu}$ at right hand side of \eqref{grad_g}.

\bibliographystyle{IEEEtran}
\bibliography{IEEEabrv}
\end{document}